# Teaching Spreadsheets: Curriculum Design Principles

Françoise Tort
STEF Laboratory – ENS de Cachan, France
francoise.tort@ens-cachan.fr

## ABSTRACT

*EuSpRIG concerns direct researchers to revisit spreadsheet education, taking into account error auditing tools, checklists, and good practices. This paper aims at elaborating principles to design a spreadsheet curriculum. It mainly focuses on two important issues. Firstly, it is necessary to establish the spreadsheet invariants to be taught, especially those concerning errors and good practices. Secondly, it is important to take into account the learners' ICT experience, and to encourage them to attitudes that foster self-learning. We suggest key principles for spreadsheet teaching, and we illustrate them with teaching guidelines.*

## 1. INTRODUCTION

EuSpRIG concern leads us to revisit spreadsheet education, taking into account errors and why they occur, auditing tools and checklists, and good practices. It requires analysing spreadsheet education according to what we know about professional users.

During the 10th annual EuSpRIG conference, E. Vandeput paved the way for a review of spreadsheet teaching [Vandeput, 2009]. In his paper, he stated two general objectives for spreadsheet education: mastering basic spreadsheet concepts, but also being able to build efficient spreadsheets. Moreover, he added that a learner should be able to learn by him/herself, discovering new functionalities according to his/her needs. We totally share this point of view that leads to elaborate three different issues:

- (i) What are the basic concepts of spreadsheets?
- (ii) How to build spreadsheets efficiently?
- (iii) What are the skills and attitudes required to be able to learn new things concerning spreadsheets by oneself.

Concerning these three issues, we also have to study what can be taught.

In this paper, we will present our personal view about these different issues. Section 2 begins with a short synthesis of what one can find concerning spreadsheet education as well as references about what should be learnt (for example, shared good practices). In section 3, we give some results about what learners know about spreadsheet. Without data concerning professional users, we will illustrate this point using our own research findings (DidaTab project). In section 4, we suggest key principles for spreadsheet teaching. In section 5, in order to illustrate our proposals, we give a non-exhaustive list of guidelines for teachers.

## 2. SPREADSHEET CURRICULUM PROLEGOMENA

### 2.1. What should one know about spreadsheet?

References about what has to be learned about spreadsheets can be found in different scientific and practical fields: ICT education, spreadsheet education, and also spreadsheet errors and good practices.

First of all: what are the basic concepts or invariants of spreadsheets? [Vandeput, 2009] uses the concept of 'organizing principle'. According to him, it refers to an automatic process carried out by a system to shorten the user's task. He elaborates an approach to help the teacher to list spreadsheet organizing principles and linked concepts.

Among the spreadsheet invariants he listed, there are basic computing concepts, like data types, operations, variables, and functions. We also suggest adding procedures, scopes of variables, data tables, sorting, etc. Indeed a spreadsheet can be seen as a program, and building spreadsheets is partly programming [Syslo and Kawientika, 2008]. This is crucial because, most of the time, spreadsheet learners have no computer science education.

To sum up, the spreadsheet teaching field includes basic programming concepts and techniques. We don't mean that it is necessary to teach computer science theory before teaching spreadsheets. We mean that it may be necessary to explain basic computing concepts during a lesson about spreadsheets, as they are the basics of spreadsheets.

Creating spreadsheets is also developing a computer-based support for a specific task. To build real and operational spreadsheets, the designer has to identify the goal of the task, the input/output data, the data processing, and translate it into spreadsheets. In 2002, Tom Grossman raised the need for the proposal of spreadsheet design processes in a spreadsheet engineering framework [Grossman, 2002]. Today, there is a substantial literature on good practices that can help designers to create efficient spreadsheets [Raffenspeger, 2001] [Bewig, 2005][O'Beirne, 2005].

Moreover, [Chadwick & Sue, 2001] claimed that "training is too frequently concentrated on 'how to do things correctly' and often ignores 'how to avoid doing things incorrectly'". According to them, one way to do so is to make spreadsheet learners more aware of common errors and to encourage them to apply checking controls during the development. Indeed, [Purser & Chadwick, 2006] show that spreadsheet error identification ability is directly affected by error-type awareness.

To sum up, it is necessary to teach the best practices to learners, in order to make them aware of the importance of spreadsheet quality.

Finally, spreadsheet programs are software tools. According to [Micheuz, 2010] software tools are ubiquitous, as they are abstract models, concrete tools and 'versatile medium' (for teaching and learning). Thus they are part of the three pillars of informatics education as depicted by Peter Hubvieser (see figure 1). An important issue is to impart practical skills combined with theoretical underpinnings. The author states the need of a model to support 'practice-theory issues', in the classroom.

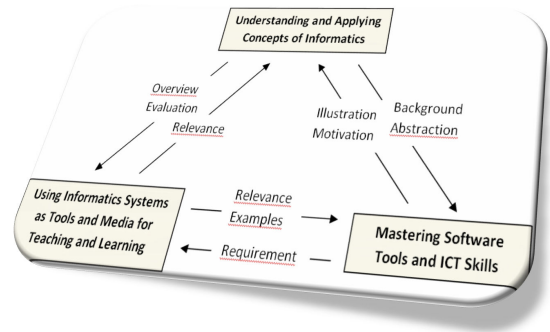

**Fig. 1: Synthesis of Informatics Education, by P. Hubvieser,
quoted in [Micheuz, 2010]**

Another important issue is what do learners really know about spreadsheets? It is important to have an idea of the gap between what we think learners should master and what they really master.

**2.2. What do learners know about spreadsheets?**

Few empirical data are available to researchers to analyze the knowledge and skills of spreadsheet users.

In the DidaTab research project, we collected data on French secondary school and undergraduate students' spreadsheet uses and skills. We can assume that the students, assessed in the project, are representative of 'new learners': novices in the use of spreadsheets, but experienced users of ICT.

Firstly, we administered computer-based tests to about 200 French students, on very basic tasks: changing cells format, writing a formula, creating a graphic or sorting data [Tort & Blondel, 2007].

The analysis of final spreadsheets provided data concerning students' productions. We found that the students succeeded in tasks dealing with superficial manipulations of spreadsheet objects that do not require specific knowledge, especially tasks linked to formating cells. All results were not as good when the tasks required more understanding of spreadsheet objects and functionalities, and with tasks concerning graphs and tables. Lastly, the more discriminating tasks where those which dealt with writing formulas [Tort et al. 2008].

In the students' responses, we even found particular errors which were sometimes surprising and difficult to classify in known taxonomies. For instance, in each assessed group, some students used functions in non-conventional manners like SUM(A1+A2+A3) or PRODUCT(A1*12,5) or SUM(A1*12,5). Some errors seem to be the symptoms of the students' misunderstanding of spreadsheet concepts, others could be the signs of confusions of students while using the spreadsheet software (see other examples in [Tort et al. 2009]).

Moreover, we collected the video records of all onscreen events during a test (for about 40 students). These videos provide useful data on the students' processes. Indeed, the analysis of screen video captures shows a large amount of details on the various ways adopted by the students when working with spreadsheets. Our findings revealed some particular modes of interaction with spreadsheet software, which could be related to the students' behaviours when using computer tools [Tort et al., 2009].

Above are some examples of observed behaviours.

**'Locked in the tool bar'.** Some students mainly used toolbar buttons (and rarely menu commands, or contextual menus, or keyboard shortcuts). They spent time searching for the buttons, and they needed several trial-errors with the buttons to achieve a task. After several trials, some of them changed their mind, lost their goal, and finally, failed in achieving the task. Because toolbar buttons activate commands with default settings, depending on the current selection, users cannot choose the parameters of the activated command.

**'Lost in the menu'.** Some students searched for a solution by systematically browsing all menus, options and buttons. They took time to read the caption of every option. And if they tried one command without obtaining success, they immediately came back to the last seen menu item or button. It looks as if students were looking for a unique command that solved the task. We assume that these students believe that every simple task has one solution and that this solution has already been scripted in the application software.

**'The copy-and-paste syndrome'.** When working on a task, some students "discovered" a command that they thought was applicable to solve a previously uncompleted task and then went backwards to the previous task and tried the 'new' command. This behaviour may be seen as a generalisation of 'copy and paste'. These students don't base their approach on a step-by-step process nor on a model of what has to be done, but on the remembering of a similar problem followed by an adaptation with trials and errors.

In addition to the specific behaviours described above, we observed, when looking at screen videos, that the mouse was always moving. The actions followed one another at a high rhythm. The students performed many gestures, with barely any pause. We wondered whether they took time to think about what they were doing.

Finally, we interviewed students just after they had performed the test to get explanations on their ways to perform tasks. Below are two interesting excerpts of the interview of a girl, age 20, enrolled in the first year of a higher national diploma (HND) in marketing.

> Q: Did you use the existing functions as SUM or AVERAGE?
> A: No, I didn't. I entered '++++' and '/'.
> Q: Do you know about these functions?
> A: Yes I do
> Q: Do you use them?
> A: No I don't because I know it makes +++ ... I'm sure to find the result
> Q: Is it faster for you?
> A: Yes it is … we don't get wrong as … as we know the formula but we can't put it into practice … so it's better to do it in one's own way … it's much easier

**Interview excerpt 1: the student explains why she prefers + operator to SUM function**

> [The student just explained she copied D14 cell formula to cells D15 to D17 using fill handle].
> Q: How do you know it provides accurate results?
> A: Well when I take it down it gives other figures... so I won't necessarily calculate after that but... it didn't give a lot of small '#' or zero ... It's OK when there are figures.

**Interview excerpt 2: the student exposes how she checks results**

The first excerpt shows that the student believes that both the + operator and the SUM function perform the same calculation. She argues her preference for the + operator by her confidence in her own way to use spreadsheets. The second excerpt shows a checking method based on the confidence in software's return (it's correct when no error value appears).

Most of the assessed students were relative novices in the use of spreadsheets, but had already personal experience of ICT. We can assume that these "new learners" have developed their own ways of working with new digital objects and computer applications. We presume that their prior experience with other software is involved in the process of spreadsheet use.

To sum up, assessed students didn't master basic spreadsheet concepts. They acquired some habits when using software, sometimes useful but most of the time not efficient. They don't base their approach on a step-by-step process. They have no idea of what spreadsheet quality means. The issue is how to get rid of such bad habits?

## 2.3. Constraints and challenges for teaching

Learners are not ICT real beginners: they commonly use ICT tools and often, they have learned by themselves. Moreover, it is likely that, most of the time, they are satisfied with their own ways of doing things. This has already been noticed by Chadwick and Sue (2001), who observe that novices overrate their own literacy skills. Learners have 'prior learning', which can be barriers to spreadsheet learning.

Learners' motivation is another point. More precisely, motivation involved by learners developing a spreadsheet is different from motivation involved in real-life situations. The impact of errors is different in an academic context and in the economic one. Learners can have difficulties in becoming aware of the role of spreadsheets and of the impact of errors in real-life situations.

Time is another constraint. Often, spreadsheet teaching takes place in short courses, sometimes dedicated to several software tools. It is not possible to show all that should be known about spreadsheets. Learners will have to continue their training beyond the course. Teachers cannot limit the class to predefined situations and go through them step by step. He/she must bring learners so that they become able to build spreadsheets in order to solve new problems, taking into account the release of up-to-date software versions.

The spreadsheet tool is also a constraint. A spreadsheet program is a piece of productivity software and its interface is designed to facilitate regular tasks and to establish a routine. According to the WYSIWYG paradigm, software interface displays results rather than underlying models. Users are encouraged to manipulate visible objects directly. When discovering new software, learners are often expecting recipes to be followed. It gives them the sensation of an immediate success ("Easy! It works!").

To sum up, teachers face a difficult issue: struggling against a deeply rooted attitude, by encouraging learners not to act as a 'push-button user', but to be curious about spreadsheet underlying concepts.

## 3. FEW KEY PRINCIPLES UNDERLYING SPREADSHEETS TEACHING

At this stage, we would like to offer key principles for spreadsheet teaching. The issue, here, is not to provide rules for selecting contents, nor organizing lessons, nor programming courses. The answer to these questions depends on the educational context: school system, grade level, training goal, etc. Our purpose is wider. We would like to offer general principles that the teacher should bear in mind when teaching spreadsheets.

### 3.1. Foster abstraction and avoid simplification

Classically, learners are shown several specific examples. By working on such diverse situations, they are supposed to be able to construct a generic model of applicability from this experience. This process of abstraction can be facilitated by some understanding of spreadsheet concepts.

On way to help learners in the abstraction process may be to explicitly use a model of spreadsheets. [Hodnigg et al. 2004] define a computational model of spreadsheets based on functional programming and data flow. [Hodnigg and Micheuz, 2008] suggest to introduce cells as 'projection-screen-devices', in order to help learners to 'see' data dependencies behind the cells' grid. Our purpose is not to discuss the model in itself but to mention it as an interesting proposal that provides an abstract framework for spreadsheet teaching. The teacher can introduce it explicitly to learners and make them work with it, or just keep it in mind when he/she explains spreadsheet principles.

Another view on abstraction may be to show learners what is common in different implementations. The same organizing principle can be implemented differently in different spreadsheet programs. Learners must be aware of implementation choices, in order to be able to focus on the concepts.

Moreover, the abstraction process relies on the quality of the examples and situations worked on in the classroom. Teachers often need to simplify situations to introduce new concepts to learners. However, they must be careful not to over-simplify examples. If learners only meet simple or specific situations (for instance, only sort one column of numbers), they will not be able to construct an effective generic model (for instance, sort a data table). Too simple examples may induce learners to apply only known procedures to new situations, by some 'copy and paste of actions'.

### 3.2. Give HCI principles instead of 'routines'

[Cziki & Zsako, 2008] suggest a typology for the teaching methods that are used in ICT education. They listed six different methods. According to them, each method is adapted to a specific educational context. The menu-oriented method is an exception, because considered as the bad one. It is based on the idea that the knowledge of an application is equal to knowing how to use an application system. It focuses on menu items, and reduces the scope of teaching to lexical knowledge and routine usage. 'It reduces the user to a dummy button-pressing machine' [Cziki & Zsako, 2008, p. 4].

We agree that, teachers should not only focus on the 'how to do'. In particular, they should very rarely write down or give a 'routine', i.e. a list of actions to perform as a recipe to be followed.

Of course routines may be necessary, as it is a guaranted way of achieving a specific goal. Some routines are more effective than others. Some effective routines result from the experience and can't be build from scratch. However, the teacher should help learners to construct routines by themselves, rather than to ask them to memorize well-known routines.

Software interfaces implement HCI (Human Computer Interface) principles, based on well-defined objects like windows, button, menu, etc, and well-defined principles like 'Toolbar buttons activate commands with default settings, depending on the current selection'. Teachers should formulate HCI principles to learners, in order to help them construct their own procedures, being aware of their choices.

### 3.3. Encourage learners to approach things critically

We met students who were overconfident with the automatic error detection by spreadsheet program (see interview excerpt 2, in previous section). More generally, a common tendency is to believe that results are correct because the computer generates them.

This overconfidence has many reasons, and among them, the lack of knowledge about errors, their nature, their origin, their impact, and how they can be (automatically) detected or not.

Without the awareness of its limit, learners can't see the computer as a tool that is under their control.

According to us the solution is not only to bring learners to master spreadsheet concepts but also to train them to approach things critically. Teachers should help them to wonder about displayed results. Teachers could also show differences between spreadsheet programs, in order to make learners understand that different implemented solutions may exist.

### 4. GUIDELINES FOR SPREADSHEET TEACHERS

Below are illustrations of how our key principles can be applied in the class. We give teaching guidelines, illustrated with examples. The list is obviously not exhaustive.

### 4.1. Explore unusual situations or uses.

It is easier to show the most common use of spreadsheet principles only. However common uses are those where results are obvious. They don't encourage the user to wonder how it works or to deepen their knowledge. It is more interesting to suggest unusual use, and to ask learners to imagine the result they would get.

**Example 1:** Once the formula copy-down principle has been discovered, suggest to copy a formula into a <u>non-adjacent</u> cell. Ask learners to imagine the resulting formula, before trying it on the computer. This is a way to get a deeper understanding of the formula copy principle.

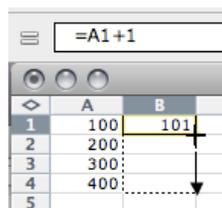 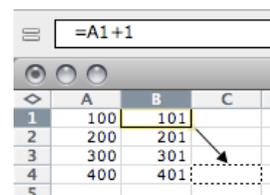

**Fig. 1a: copying formula, a common situation**　　**Fig 1b: copying formula, an unusual use**

**Example 2**: Once data sorting is known, suggest to the learners to sort alphanumeric values (starting with numbers). Ask them to imagine the result. This is a way to get a deeper understanding of the different data types and their main characteristics.

**Fig 2a: learners' prevalent idea on address sorting**   **Fig2b: result of address sorting**

**4.2. Explore implementation choices and how to customize options.**

A simple way to make learners aware of implementation choices is to show them differences between different spreadsheet programs.

**Example 1:** Try to auto-fill '1998-1999' in both Open Office Calc and Microsoft Excel. Ask learners to wonder why it is different, how it works in both programs. Explain that the same feature is differently implemented.

**Fig 3a: Auto-fill of '1998-1999'**          **Fig 3b: Auto-fill of '1998-1999'**
**with Excel Microsoft 2003**                 **with Open Office Calc (3.2.0)**

**Example 2:** Try to auto-fill 'Monday', 'January', and 'One'. Why the auto-fill of 'one' does not give 'two-three-four'? Show how to customize auto fill lists.

**Fig 4a: Auto-fill of 'Monday'**             **Fig 4b: Auto-fill of 'One'**

**Example 3:** Date format provides many examples. For instance, the option 'keep years typed on 4 digits' in general preferences. When it is unchecked, entering '10/10/2010' gives '10/10/10'.

**4.3. Explore spreadsheet limits in error detection**

It is important to explain which errors spreadsheet programs can detect. The errors automatically detected are: syntax errors (for instance, errors in the name of formula), errors in data type, impossible results (especially in lookup functions). Instead, spreadsheet program can't check whether a calculation outcome is efficient.

**Example 1:** Users may be misled about automatic error detection capability. Especially, when an undetectable error causes a detected one, for which an error value occurs. The user may believe that the spreadsheet detects the first one (see figure 5). When learners meet such situations, ask them which error is detected and show them a similar case in which no error message occurs.

**Fig 5a and b: The formula '=B4*F1' was entered in C4 and copied down to C5:C7. The right formula would have been '=B4*$F$1'. The program can't detect such error. In the first case (Fig. a), error values occur, whereas no error occur in second case (Fig. b).**

**Example 2:** Showing an incorrect solution proposed by the spreadsheet autocorrection feature. [Walkenbach, 2007] gives an interesting example. It occurs with Microsoft Excel 2007, when the user forgets a parenthesis inside a formula, and the autocorrection proposal is the addition of a parenthesis at the end of the formula.

**Fig. 6: The user enters '=AVERAGE(SUM(A1:A4;SUM(A5:A7))', forgetting a parenthesis after A4. Microsoft Excel 2007 autocorrection adds the parenthesis at the end of the formula.**

### 4.4. Explore borderline uses

Tutorials and help software features always describe the good manner to use spreadsheets. Often, teachers only show the best solutions to exercices. However, exploring the borderline uses, in which it doesn't (or shouldn't) work, may help understanding underlying concepts.

**Example:** When discovering a new function, try it with different data types than those expected. Observe and interpret the outputs.

**Fig. 7: use of operator * with different data types as operands.**

## 4.5. Explore spreadsheet quality and ways to improve

The quality of the spreadsheet is related to its reliability, understandability, testability and extendibility [McKeever et al., 2009]. Spreadsheet quality can be enhanced applying the best practices during spreadsheets design. Some good practices can be easily understood and mastered by learners; others may be more difficult to appreciate.

**Example 1:** Check situations when the meaning of data is ambiguous because the spreadsheet use and issue is not clear. Discuss the issue of data representation and calculus.

For instance, giving the spreadsheet on figure 8, the learner is asked to calculate the number of months remaining before the end of the year. There are at least two answers: one calculates the number of empty cells in the range B4:B15, the other calculates the difference between the current month's range and the range of december (12). The issue here is not to write the formula. The question is: are both calculi equivalent?

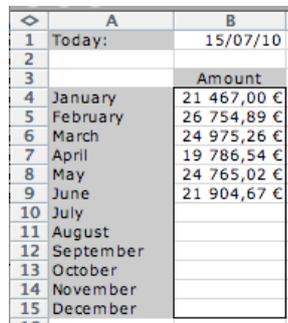

**Fig. 8: How to calculate the number of month remaining before the end of the year?**

**Example 2:** Give a spreadsheet containing errors. Ask learners to fix them. Explore spreadsheet audit tools.

- Typographical error in data (accidental error)
- Cells containing space characters instead of being empty
- Relative cell references instead of absolute cell references
- Hard-coding of a constant in a formula
- One different formula in an array of identical formula (structural error).

**Fig. 9: Examples of interesting errors to fix**

**Example 3:** Introduce some of the best practices. Some good practices require to master more advanced concepts and may be introduced later in the class (see the last two items in the list of figure 10).

- Put visible label on data spreadsheets and use the label as cell names
- Ordering and grouping data into areas, according to their use and meaning.
- In order to avoid hard coding of constants in formulas, put constant data in dedicated areas and use (absolute) cell references in formulas.
- Divide calculus in intermediate formulas, entered in different cells.

> - In order to give a code to values according to a mapping between intervals and codes, use lookup function rather than 'nested if'.
> - Prefer array formulas rather than copying-down formula into cell array.

**Fig. 10: Examples of interesting good practices**

## 5. DISCUSSION

There is a positive gap between the learners' spreadsheet mastering and what we think they should master. It is likely that their self-experience on ICT tools encourages them to be satisfied with 'surface knowledge'. They may not be curious about underlying concepts. Teachers don't only need to explain underlying concepts, to train learners to spreadsheet programs use, but also to encourage attitudes that foster self-learning.

In the paper, we offered key principles for spreadsheet teaching and illustrated them with some guidelines. According to us, important issues are open.

Firstly, studying spreadsheet professional users' practices and actual knowledge and skills remains an important issue. Spreadsheet error research focuses on the analysis of users' final spreadsheets. They provide data on users' results, not on their process of spreadsheet use. In the field of ICT research, many studies investigate users competencies, by the mean of surveys. They are mostly based on self-assessment techniques, for which the major drawback is that users largely over-estimated their own skills. It would be interesting to get empirical data on the ways professional users design spreadsheets.

Secondly, another issue concerns good practices teaching. The literature on the subject is substantial. Often, papers formulate checklists or catalogues of good practices. They are numerous and concern many aspects of spreadsheets: data presentation, data display, calculation, error, specification, design, testing, etc. [O'Beirne, 2005] offers an interesting way of organizing good practices in categories and skill sets. It is interesting to carry on this effort. In particular, it would be necessary to describe good practices, regarding spreadsheet concepts and skills involved. Indeed, some good practices are based on advanced concepts and uses of spreadsheets, and are more specialized. On the contrary, there are basic good practices that should be known even by spreadsheet novices.